\documentclass[11pt]{amsart}
\usepackage{amsmath,amsthm,amssymb,amscd, graphicx, wrapfig}
\usepackage[all]{xy}
\usepackage[margin=1.1in]{geometry}
\usepackage[usenames, dvipsnames]{color}

\usepackage[T1]{fontenc}

\theoremstyle{plain}
\newtheorem{thm}{Theorem}

\newtheorem{lem}[thm]{Lemma}
\newtheorem*{conj}{Conjecture}

\theoremstyle{definition}
\newtheorem*{defn}{Definition}

\newtheorem*{exmp}{Example}

\theoremstyle{remark}
\newtheorem*{rem}{Remark}

\newtheorem*{ack}{Acknowledgments}

\numberwithin{equation}{section}

{\makeatletter
 \gdef\xxxmark{%
   \protected@write\@auxout{\def\PAGE{ page }}
     {\@percentchar xxx: section \thesubsubsection \PAGE \thepage}%
   \expandafter\ifx\csname @mpargs\endcsname\relax 
     \expandafter\ifx\csname @captype\endcsname\relax 
       \marginpar{xxx}
     \else
 xxx 
     \fi
   \else
 xxx 
   \fi}
 \gdef\xxx{\@ifnextchar[\xxx@lab\xxx@nolab}
 \long\gdef\xxx@lab[#1]#2{{\bf [\xxxmark #2 ---{\sc #1}]}} 
 \long\gdef\xxx@nolab#1{{\bf [\xxxmark #1]}}
 \gdef\turnoffxxx{\long\gdef\xxx@lab[##1]##2{}\long\gdef\xxx@nolab##1{}}%
}

\newcommand{\R} {{\mathbb R}}                              

\newcommand{\lst} {\mathcal L}

\newcommand{\hide}[1]{}

\begin{document}

\title{Unfoldings and Nets of Regular Polytopes}

\author{Satyan L.\ Devadoss}
\address{S.\ Devadoss: University of San Diego, San Diego, CA 92110}
\email{devadoss@sandiego.edu}

\author{Matthew Harvey}
\address{M.\ Harvey: The University of Virginia's College at Wise, Wise, VA 24293}
\email{msh3e@uvawise.edu}

\begin{abstract}
Over a decade ago, it was shown that every edge unfolding of the Platonic solids was without self-overlap, yielding a valid net.  We consider this property for regular polytopes in arbitrary dimensions, notably the simplex, cube, and orthoplex. It was recently proven that all unfoldings of the $n$-cube yield nets. We show this is also true for the $n$-simplex and the $4$-orthoplex but demonstrate its surprising failure for any orthoplex of higher dimension.
\end{abstract}


\maketitle
\baselineskip=17pt

%
%
\section{Introduction} \label{s:intro}

The study of unfolding polyhedra was popularized by Albrecht D\"urer  in the early 16th century who first recorded examples of polyhedral \emph{nets}, connected edge unfoldings of polyhedra that lay flat on the plane without overlap. Motivated by this, Shephard  \cite{sh1} conjectures that every convex polyhedron can be cut along certain edges to admit a net.  This claim remains tantalizingly open and has resulted in numerous areas of exploration; see \cite{gho} for a survey.

We consider this question for higher-dimensional \emph{polytopes}: The codimension-one faces of a polytope are  \emph{facets} and its codimension-two faces are \emph{ridges}. The analog of an edge unfolding of polyhedron is the \emph{ridge unfolding} of an $n$-dimensional polytope: the process of cutting the polytope along a collection of its ridges so that the resulting (connected) arrangement of its facets develops isometrically into an $\R^{n-1}$ hyperplane.
There is a rich history of higher-dimensional unfoldings of polytopes, with the collected works of Alexandrov \cite{alex1} serving as seminal reading.  In particular, Buekenhout and Parker \cite{bupa} enumerate the ridge unfoldings of the six regular convex 4-polytopes.  

In our work, instead of trying to find one valid net for each convex polyhedron (as posed by Shephard), we consider a more aggressive property:

\begin{defn}
A polytope $P$ is \emph{all-net} if every ridge unfolding of $P$ yields a valid net.\footnote{This nomenclature comes from Joe O'Rourke: akin to a basketball ``all-net'' shot that scores by not touching the rim, all unfoldings become successful nets by facets not overlapping and touching each other.}
\end{defn}

\begin{figure}[h]
\includegraphics[width=.95\textwidth]{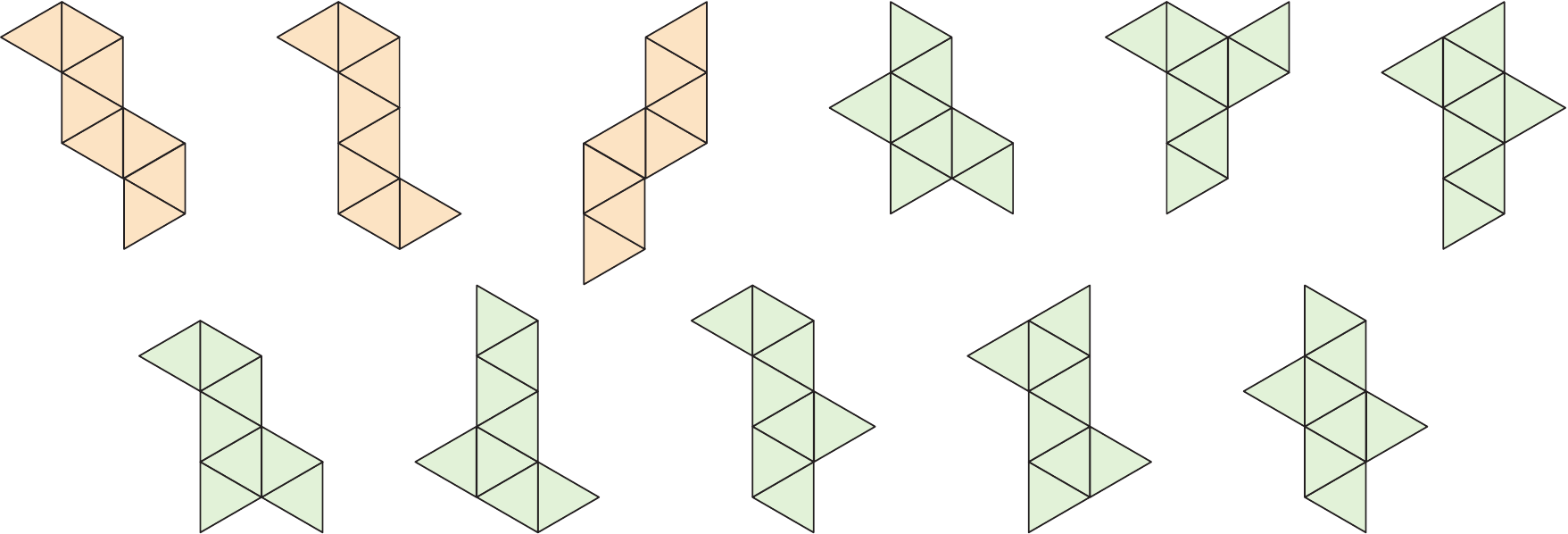}
\caption{The 11 nets of the octahedron, also known as the 3-orthoplex.}
\label{f:3dnet}
\end{figure}

\noindent A decade ago, Horiyama and Shoji \cite{hosh} showed that the five Platonic solids are all-net. This is easy to check for the tetrahedron, cube, and octahedron, for which there are few unfoldings, but it is significantly nontrivial for the dodecahedron and icosahedron, for which there are 43,380 distinct unfoldings. Figure~\ref{f:3dnet} shows the 11 different unfoldings (up to symmetry) of the octahedron, all of which are nets.  Recent work \cite{pnas} has shown applications in protein science: polyhedral nets are used to find a balance between entropy loss and energy gain for the folding propensity of a given shape.

The higher-dimensional analogs of the Platonic solids are the regular polytopes. Three classes of regular polytopes exist for all dimensions: $n$-simplex, $n$-cube, and $n$-orthoplex (sometimes called the \emph{cross-polytope}). Three additional regular polytopes appear only in four-dimensions: the 24-cell, 120-cell, and 600-cell. 
It was recently shown that the $n$-cube is all-net \cite{ddrw}; in particular, the 261 different ridge unfoldings of the $4$-cube do not overlap.  

In Section~\ref{s:simplex}, we show that the $n$-simplex is all-net  as well.  Section~\ref{s:orthoplex} establishes combinatorial and geometric attributes of the orthoplex which Section~\ref{s:orthonet} uses to demonstrate the all-net property of the $4$-orthoplex.  That is, the 110,912 ridge unfoldings of the 4-orthoplex enumerated in \cite{bupa} are without overlap.  Surprisingly, for all $n >4$, the $n$-orthoplex fails to be all-net; we provide explicit counterexamples to a conjecture in \cite{ddrw}.  Thus only three regular polytopes stand with unresolved all-net traits: the 24-cell, 120-cell, and 600-cell.  

\begin{rem}
We encourage the reader to explore a lovely interactive software \cite{zhang} by Sam Zhang  that creates every net of the 4-cube, 4-simplex, and 4-orthoplex by drawing  on its dual 1-skeleton.
\end{rem}

\begin{ack}
We thank Nick Bail, Zihan Miao, Andy Nelson, and Joe O'Rourke for helpful conversations. 
The first author was partially supported from an endowment by the Fletcher Jones Foundation.
\end{ack}

%
%
\section{Unfolding the Simplex} \label{s:simplex}
\subsection{Lists and Chains}

We explore ridge unfoldings of a convex polytope $P$ by focusing on the combinatorics of the arrangement of its facets in the unfolding. In particular, a ridge unfolding induces a spanning tree in the 1-skeleton of the dual of $P$ : a tree whose nodes are the facets of the polytope and whose edges are the uncut ridges between the facets \cite{sh2}.
Our focus throughout this paper will be on the $n$-simplex and the $n$-orthoplex, both of whose facets are $(n-1)$-simplices. We begin by studying  paths in the 1-skeleton, corresponding to a chain of unfolded simplicial facets.

\begin{defn}
A \emph{list} $\lst=\langle a_1, a_2, \dots, a_k \rangle$ is a sequence of numbers from $\{1,\ldots, n\}$ (possibly with repeats) where with no number is listed twice in a row. 
\end{defn}

\begin{figure}[h]
\includegraphics[width=.9\textwidth]{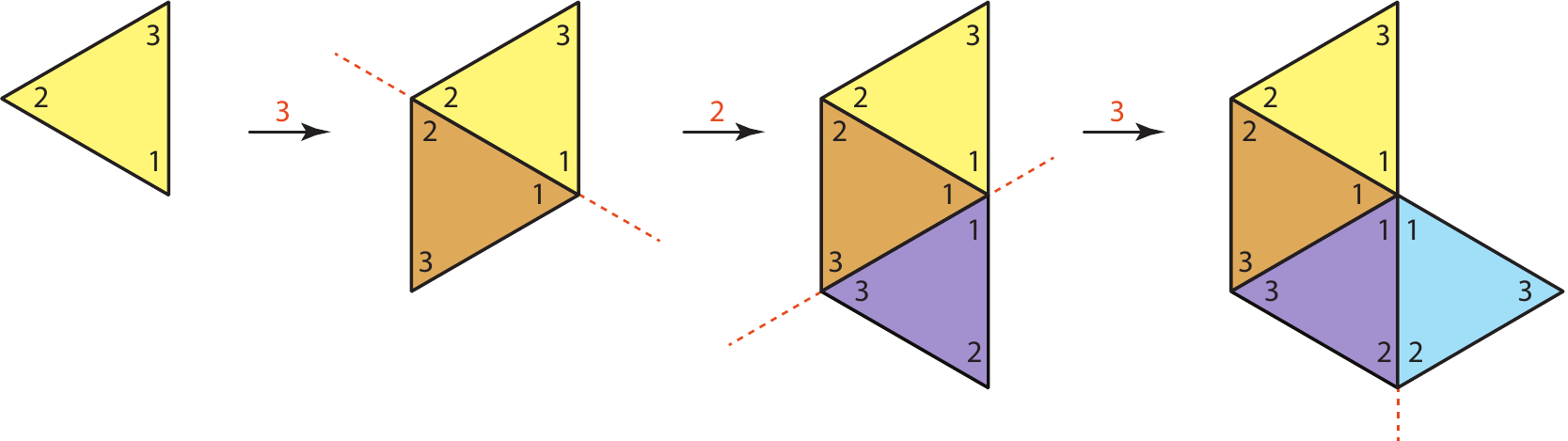}
\caption{The chain of simplices assembled from the list $\langle 3,2,3\rangle$.}
\label{f:lists}
\end{figure}

Label the vertices of the $(n-1)$-simplex $S$ with the numbers $1,\ldots, n$.  Given a list $\lst$ with $k$ elements, we construct a chain $C(\lst$) of $k+1$ simplices from the list as follows: Starting with $S = S_1$, attach a simplex $S_2$ to $S_1$ on the facet of $S_1$ that is opposite vertex $a_1$.  Note that all but one of the vertices of $S_2$ will inherit a label from $S_1$ and we label the remaining one $a_1$.  Attach a third simplex $S_3$ to $S_2$ on the facet opposite vertex $a_2$, and extend the labeling from $S_2$ to $S_3$ as before, and continue in this matter until the list is exhausted.  Figure~\ref{f:lists} shows this process in action for the list $\langle 3,2,3 \rangle$, creating a chain of four 2-simplices.

\subsection{Matrix Coordinates}

\begin{figure}[h]
\includegraphics[width=.9\textwidth]{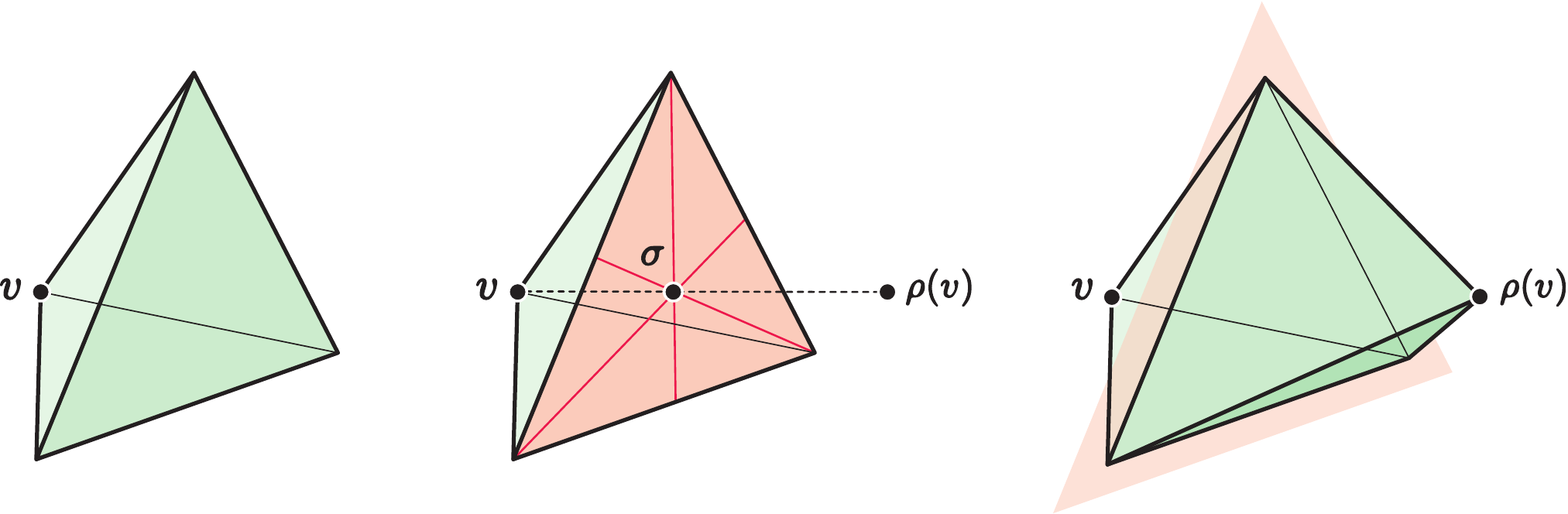}
\caption{The reflection of the vertex across the opposite face.}
\label{f:reflection}
\end{figure}

Beyond this combinatorial formation, we introduce a coordinate system to capture the geometry. Begin by placing the $n$ vertices of the $(n-1)$-simplex $S$ at the standard basis vectors $e_i$ of $\R^n$.  Note that the coordinates of its vertices are recorded as the column vectors of the $n\times n$ identity matrix. The rest of the chain is then placed in the hyperplane $x_1 + \cdots +x_n=1$ by a sequence of reflections.
Let $\rho$ denote the reflection of $S$ across its facet opposite the vertex (say $v$) labeled with number $a_1$.  Thus, $\rho$ fixes all vertices except for $v$; see Figure \ref{f:reflection}. 

To calculate the coordinate of $\rho(v)$ in $\R^n$, we first find the center $\sigma$ of the facet opposite $v$:
    $$\sigma =\left (\frac{1}{n-1}, \ldots,  \ 0,\  \ldots, \frac{1}{n-1} \right ),$$
where  $0$ occurs in the $a_1$-{th} coordinate. Since $\sigma$ bisects the segment from $v$ to $\rho(v)$,
    \begin{align*}  
    \rho(v) & = \ v + 2 \ \overrightarrow{v \, \rho(v)} \\ 
    & = \ \left ( 0, \ldots, 1, \ldots 0 \right ) \ + \ 2 \left ( \frac{1}{n-1},  \ldots, -1, \ldots, \frac{1}{n-1} \right ) \\
		&= \ \left ( \frac{2}{n-1}, \ldots, -1, \ldots, \frac{2}{n-1} \right ),
	\end{align*}
where the $-1$ occurs in the $a_1$-{th} coordinate. 
Hence the reflection $\rho$ is given by a matrix $M_{a_1}$, which is the identity except for $\rho(v)$ in the $a_1$-{th} column. 
Thus the coordinates of the  $i$-th vertex of $S_2$ are recorded in the $i$-th column $v_i$ of $N_1 = M_{a_1}$. By change of coordinates, its image under the reflection from $S_2$ to $S_3$ is 
     \[ N_1 M_{a_2} N_1^{-1} v_i = N_1 M_{a_2} e_i \,,\]
and thus, the coordinates of the $i$-th vertex of $S_3$ are recorded in the $i$-th column of $N_2 = N_1 M_{a_2}$. Note that because $M_{a_2}$ affects only the $a_2$ column, $N_1$ and $N_2$ differ only in the $a_2$ column. Continuing in this way, the vertices of $S_{k+1}$ are recorded as the columns of $N_k = N_{k-1} M_{a_k}$.

\begin{exmp} For two-dimensional chains (where $n=3$), the reflection matrices are  
  \[ M_1 = \begin{bmatrix} -1 & 0 & 0 \\ 1 & 1 & 0 \\ 1 & 0 & 1 \end{bmatrix}, \quad  M_2 = \begin{bmatrix} 1 & 1 & 0 \\ 0 & -1 & 0 \\ 0 & 1 & 1 \end{bmatrix}, \quad  M_3 =\begin{bmatrix} 1 & 0 & 1 \\ 0 & 1 & 1 \\ 0 & 0 & -1 \end{bmatrix}.  \]
For the chain generated by $L=\langle 3,2,3 \rangle$, the coordinates of the vertices of the second, third, and fourth simplices are the column vectors of the matrices
 \[N_1 = M_3 = \begin{bmatrix} 1 & 0 & 1 \\ 0 & 1 & 1 \\ 0 & 0 & -1 \end{bmatrix}, \quad  
  N_2 = N_1M_2= \begin{bmatrix} 1 & 2 & 1 \\ 0 & 0 & 1 \\ 0 & -1 & -1 \end{bmatrix}, \quad  
  N_3 = N_2M_3 = \begin{bmatrix} 1 & 2 & 2 \\ 0 & 0 & -1 \\ 0 & -1 & 0 \end{bmatrix}. \]
We remark that these are the barycentric coordinates of the vertices relative to the first simplex. 
\end{exmp}

\subsection{Simplex Unfolding}

An $n$-simplex has $n+1$ facets, and each is adjacent to every other. Thus, any listing of the facets (without repeat) describes a chain. However, because the full symmetric group acts transitively on the simplex, there is only one possibility:

\begin{lem} \label{l:congruent}
A chain of length $k+1$ in an unfolding of an $n$-simplex is congruent to $C(\langle 1,2,\ldots, k\rangle)$.   
\end{lem}

\hide{
\begin{proof}
Let $C$ be a chain with length $k+1$. Number the facets of the $n$-simplex: $0$ for the first facet in the chain, $1$ for the second, and so on. If $k<n$, the remaining facets can be numbered arbitrarily. Then label the vertices so that vertex $i$ is opposite facet $i$. Those vertex labels descend to the unfolded chain. They agree with the vertex labeling of the $0^{th}$ facet of $C(\langle 1\ldots, k \rangle)$, however not necessarily with subsequent facets, as shown in fig ..?.. \xxx{S: I think we probably do need a picture here to show what is going on.} The process of of unfolding the chain and the process of attaching facets to form $C(\langle 1,\ldots,k\rangle)$ are the same as long as the labels on the vertices that govern the procedure agree. Since each number in the chain occurs only once, the only vertices that are referenced are in the $0^{th}$ facet, so the two chains are the same. 
\end{proof}
}

When $C(\langle 1,\ldots,k\rangle)$ is embedded in $\R^n$ as described above, the coordinates of the vertices of its facets are the columns of the matrices $N_i = N_{i-1} M_i$, where $N_0=I$.  Furthermore, the points in the interior of a facet are linear combinations of its vertex coordinates,  $c_1 v_1 + \cdots + c_n v_n$, where each $c_i$ is positive, and $c_1+\cdots +c_n=1$. Therefore, to see whether this chain overlaps itself, we need to look carefully at the values in those matrices.

\begin{lem} \label{l:positive}
The first row of $N_k$ contains no positive terms when $k>0$. 
\end{lem}

\begin{proof}
We prove the more precise statement that the $(1,j)$-entry is negative if $j\le k$ and zero if $j>k$.  The proof is by induction and the base case is apparent. Assuming the result for $N_{l-1}$, the product $N_l=N_{l-1} M_l$ has all the same entries as $N_{l-1}$ except in the $l$ column. Hence the $(1,j)$-entries remain negative when $j<l$ and zero when $j>l$. If $a_{ij}$ denotes the $(i,j)$-entry of $N_{l-1}$, then the $(1,l)$-entry of $N_l$ (from the dot product of row 1 of $N_{l-1}$ and column $l$ of $M_l$) is given by
$$ \bigg( a_{1,1}, \ \ldots,  \ a_{1,l-1}, \ 0, \ \ldots, \ 0 \bigg) \ \cdot  \ \left ( \frac{2}{n-1}, \ \ldots, \ -1, \ \ldots, \ \frac{2}{n-1} \right ),
$$
where the $-1$ occurs in the $l$ coordinate. This becomes
$$ a_{1,1} \  \bigg(\frac{2}{n-1}\bigg)  \ + \ \cdots \ + \  a_{1,l-1} \  \bigg(\frac{2}{n-1}\bigg) \ + \ 0 \ (-1) \ + \ 0 \  \bigg(\frac{2}{n-1}\bigg) \  + \ \cdots \ + \ 0 \  \bigg(\frac{2}{n-1}\bigg) \,, $$
which is clearly negative.
\end{proof}

\begin{thm}
Every unfolding of the $n$-simplex yields a net.
\end{thm}

\begin{proof}
Assume otherwise, where there exists an unfolding of the simplex with two overlapping simplicial facets.  The path (in dual 1-skeleton) connecting these two facets forms a chain of simplices. By Lemma~\ref{l:congruent}, that chain is congruent to $C(\langle 1, 2,\ldots, k \rangle)$ for some $k$, where facet $f_0$ and facet $f_k$ must overlap. By construction, the interior points of facet $f_0$ have all positive coordinates.  For facet $f_k$, Lemma~\ref{l:positive} guarantees that there are no positive entries in the first row; hence, all interior points of $f_k$ must have a negative first coordinate. Thus, the two facets cannot intersect, resulting in a contradiction.
\end{proof}

%
%
\section{Orthoplex Combinatorics and Geometry}  \label{s:orthoplex}
\subsection{Valid Lists}

In contrast to the simplex, both unfoldings of the $n$-orthoplex and the chains within these unfoldings exhibit considerable variety. Unfoldings of the $n$-orthoplex are in bijection with spanning trees of the 1-skeleton of the $n$-cube.  Consider the following approach to record paths on this skeletal structure:  Position the $n$-cube with antipodal vertices at $(0, \dots, 0)$ and $(1, \dots, 1)$. A path along the edges of this cube is encoded as a list of binary numbers  where exactly one digit changes from one entry to the next.\footnote{This list of binary numbers is called a \emph{Gray code}.}  For our work, our list $\lst$ simply records the digit entry that changes in moving from one vertex to another. Thus by duality, the ridges of the orthoplex inherit these labels and the process of unfolding the chain corresponds to the construction of $C(\lst)$.

\begin{figure}[h]
\includegraphics[width=.9\textwidth]{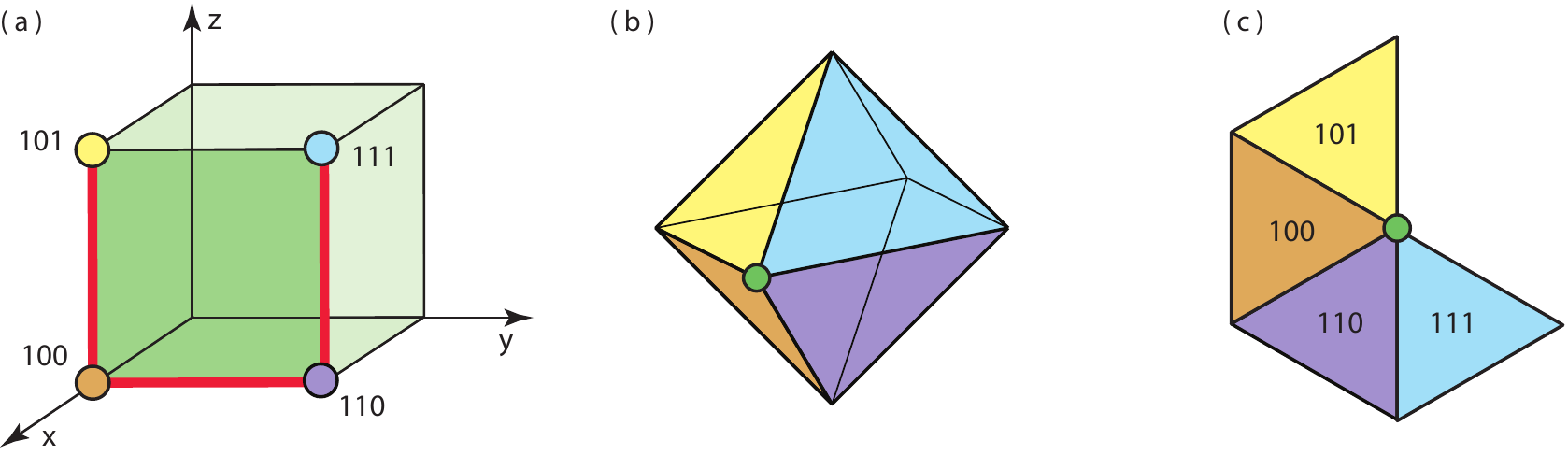}
\caption{Path on the 3-cube and a partial unfolding of the octahedron.}
\label{f:codes}
\end{figure}

\begin{exmp}
Consider the Gray code $\langle 101, 100, 110, 111\rangle$ associated to the  list $\langle 3, 2, 3\rangle$.   Figure~\ref{f:codes}(a) shows the path on four vertices of the cube, (b) corresponding to four adjacent facets of the octahedron, (c) resulting in a partial chain unfolding.  Compare to Figure~\ref{f:lists}.
\end{exmp}

\begin{rem}
Up to symmetry, there are just three spanning paths on the 1-skeleton of the $3$-cube: $\langle 1, 2, 1, 3, 1, 2, 1\rangle$, $\langle 1,2,1,3,2,1,2 \rangle$, and $\langle 1,2,3,2,1,2,3\rangle$, corresponding to the first three highlighted nets shown in Figure~\ref{f:3dnet}. The situation escalates rapidly as $n$ increases: there are 238 spanning paths on the $4$-cube and 48,828,036 on the $5$-cube \cite{oeis}.
\end{rem}

\begin{defn}
A list of numbers from $\{1, \dots, n\}$ is \emph{valid} if it corresponds to a path on the $n$-cube.
\end{defn}

\noindent Note that every path naturally yields a valid list, and given the starting position of the path, every valid list yields a path.  The following provides a combinatorial check for list validity:

\begin{lem}\label{lem:validCheck}
A list is valid if and only if it contains no sublist of consecutive entries in which each entry occurs an even number of times. 
\end{lem}

\begin{proof}
For this proof, we position the $n$-cube with antipodal vertices at $(-1, \dots, -1)$ and $(1, \dots, 1)$. Consider a list 
$\lst  = \langle a_1, \cdots, a_k\rangle$
and label the vertices of the path it generates as $v_1, \ldots, v_{k+1}$, starting from $v_1=(1,1,\ldots,1)$. Then $v_{i+1}$ is the image of $v_i$ under the reflection across the hyperplane $x_{a_i}=0$. This reflection is given by the $n \times n$ diagonal matrix $M_i$ whose $(i,i)$-entry is $-1$, and whose remaining diagonal entries are $1$. Suppose $\lst$ contains a sublist $a_j, a_{j+1}, \ldots, a_l$  in which  all entries occur an even number of times. Then, because each $M_i$ is its own inverse and they all commute, 
 $$M_l  \cdot M_{l-1}  \cdots  M_j \cdot v_{j-1} \ = \ I \cdot v_{j-1} \ = \ v_{j-1} \,.$$
Thus, the path is truly a loop and the list is invalid.

Conversely, if the sequence describes a loop, so that $v_j=v_k$ for some $j \ne k$, then $v_j$ is a fixed point of 
$$N \ = \ M_{l-1} \cdot M_{l-2} \cdot \ \cdots \ \cdot \ M_{j+1}\, .$$
Since $N$ is a diagonal matrix, it cannot have fixed points of the form $(\pm 1, \pm 1, \ldots, \pm 1)$ unless $N$ is the identity matrix.  Thus, $M_i$ must occur an even number of times in the product. 
\end{proof}

\begin{rem}
With the characterization given by this lemma, it is straightforward to create an algorithm to build valid lists: recursively append numbers $\{1,\dots, n\}$ and check whether any of the new consecutive sublists have entries that occur an even number of times.
\end{rem}

\subsection{Centroids}

The question of whether two facets overlap depends on how close they are to each other, which can be estimated by calculating the distance between their centroids.  If the vertices are $v_i = (a_{i1}, \ldots a_{in})$, the centroid is found by averaging their coordinates: 
$$\left (\ \frac{1}{n}\sum a_{1j}, \ \ldots,  \ \frac{1}{n}\sum a_{nj} \ \right ) \,.$$

\begin{lem} \label{l:centroid}
Let $d$ denote the distance between the centroids of two $(n-1)$-simplex facets of the $n$-orthoplex in an unfolding. If $d<2/\sqrt{n(n-1)}$, the facets must intersect. If $d>2\sqrt{(n-1)/n}$, the facets cannot intersect.	
\end{lem}
 
\begin{proof}
The closest point to the centroid on the surface of the facet is the midpoint of a face. Since all the facets are identical, we use the initially placed facet to compute the distance: The centroid of the $(n-1)$-simplex is at $(1/n,\ldots,1/n)$ and the center of one of the faces is at 
$$(1/(n-1), \ \ldots, \ 1/(n-1), \ 0) \,.$$ 
The distance between them is 
	\[ \sqrt{\left (\frac{1}{n}-\frac{1}{n-1}\right )^2(n-1)+\left (\frac{1}{n}-0\right )^2}
	=\frac{1}{\sqrt{n(n-1)}} \ .\]
If centroids are separated by less than double that amount, the facets must overlap. 

The farthest point from the centroid to the surface of the facet is a vertex, one of which is $(0,\ldots,0,1)$. The distance between them is
	\[ \sqrt{ \left ( \frac{1}{n}-0 \right )^2(n-1) + \left ( \frac{1}{n}-1 \right )^2} 
	= \sqrt{\frac{n-1}{n}} \ . \]
If centroids are separated by more than double that amount, the facets cannot overlap.	
\end{proof}

%
%
\section{Orthoplex Unfolding}  \label{s:orthonet}
\subsection{Path Extensions}

\begin{figure}[h]
\includegraphics[width=.9\textwidth]{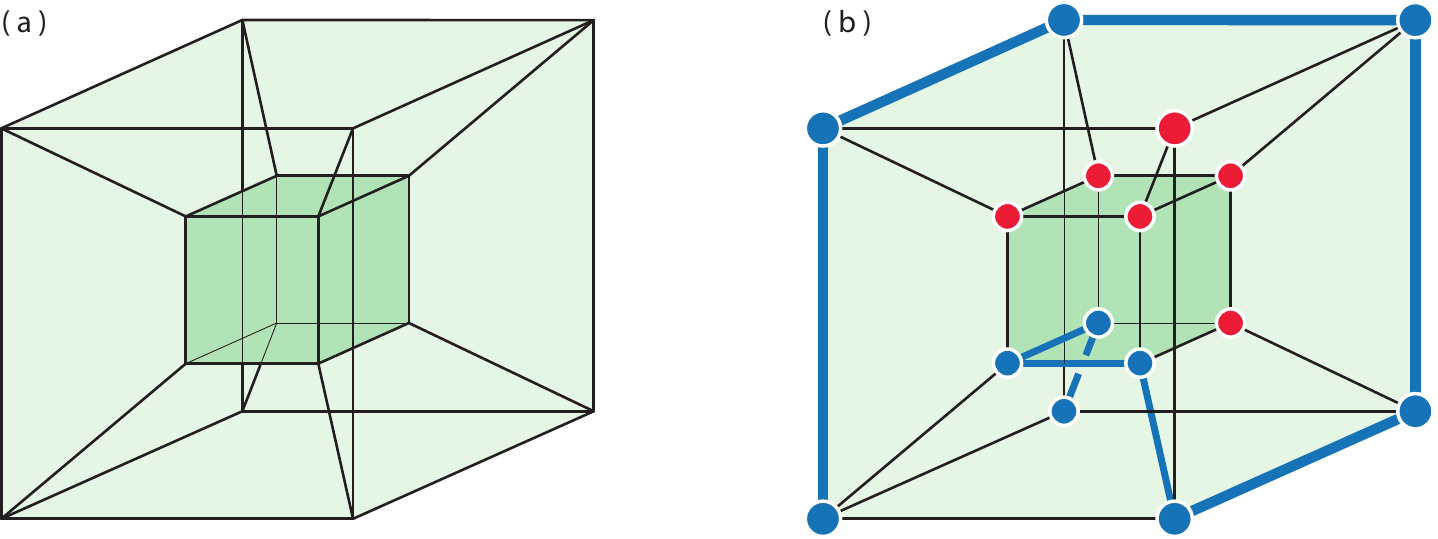}
\caption{A path in the 4-cube that cannot be further extended.}
\label{f:4-path}
\end{figure}

This section proves that the $4$-orthoplex is all-net.  We do this by extending paths on the skeleton of the $4$-cube.  While any path along a $3$-cube can always be extended to a spanning path, this is not true for $n \geq 4$.  For example, Figure~\ref{f:4-path}(a) shows the 1-skeleton of the 4-cube, and the blue path shown in (b) cannot be extended further.   However, there is a partial result.

\begin{lem} \label{l:8less}
A path on skeleton of the $4$-cube can be extended to connect at least nine vertices.
\end{lem}

\begin{proof} Let $s$ be the starting point and $e$ the ending point of a path that cannot be extended further. Let $V_s$ be the set of the four vertices adjacent to $s$, and $V_e$ be the set of the four vertices adjacent to $e$. Since the path cannot be extended, it must already pass through all the vertices in $V_e$ and $V_s$. If $s$, $e$, and the vertices of $V_s$ and $V_e$ are all distinct, then the path connects at least ten vertices. Even if a pair of listed vertices coincide, the path connects at least nine. However, there are two ways in which it is possible for $\{s\} \cup \{e\} \cup V_s \cup V_e$ to contain only eight vertices.

First, if $s$ and $e$ are adjacent, then $e \in V_s$ and $s \in V_e$. In this case, without loss of generality, we may assume the path starts at $(0, 0, 0, 0)$ and ends at $(1, 0, 0, 0)$. Then 
\begin{align*} 
V_e &= \{(0,1,0,0), (0,0,1,0), (0,0,0,1)\}, \\
V_s &= \{(1,1,0,0), (1,0,1,0), (1,0,0,1)\}.  
\end{align*}
However, a simple check shows that there is no path containing only these eight points. In fact, no such path can contain more than four of them.

Second, if $s$ and $e$ are opposite corners of a codimension two (square) face, then $V_s$ and $V_e$ share two vertices. Color a vertex red if the sum of its coordinates is even, and blue if it is odd. Any path connecting $s$ and $e$ will pass through nodes that alternate in color. Since $s$ and $e$ are opposite corners of a square, they must be the same color (say, red). Then all six vertices of $V_s$ and $V_e$ must be blue. Thus, a path that passes through all six of them must also pass through at least five red vertices, yielding a path with length greater than nine.
\end{proof}

\subsection{Unfolding the $4$-orthoplex}

Rephrasing Lemma~\ref{l:8less} differently, we can say that any valid list can be extended to a valid list with at least eight entries.  We use this to begin unfolding the $4$-orthoplex.

\begin{lem} \label{l:eight} 
Every valid list containing exactly eight entries unfolds to form a partial net of the $4$-orthoplex.	
\end{lem}

\begin{figure}[h]
\includegraphics[width=.5\textwidth]{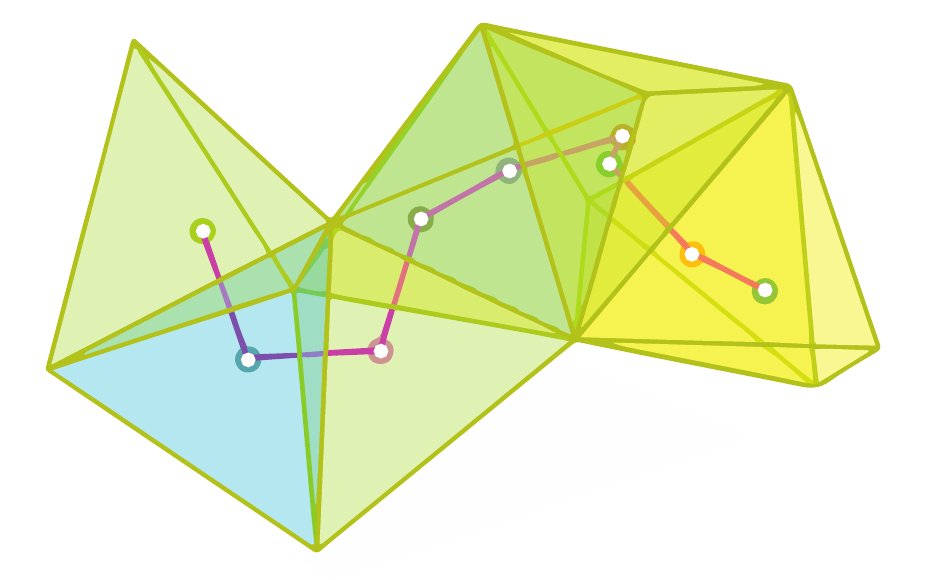}
\caption{The partial unfolding corresponding to a valid list with length eight.}
\label{f:8length}
\end{figure}

\begin{proof}
Using the check for valid lists described in Lemma~\ref{lem:validCheck}, there are 128 valid lists. Taking advantage of reverse symmetry\footnote{The unfolding given by $\langle 1,2,1,3\rangle$ is congruent to that given by $\langle 3,1,2,1\rangle$, which can be renumbered $\langle 1,2,3,2\rangle$.} 
reduces the number to 66 (where four lists are their own reverses). One such is shown in Figure~\ref{f:8length}. By direct inspection, none of them self-intersect.
\end{proof}

\begin{lem}\label{l:8more}
If two facets of the $4$-orthoplex are separated by eight or more facets, they cannot overlap.         
\end{lem}

\begin{proof}
This is verified computationally by generating all valid lists with length nine or more facets and then computing distances between centroids of facets that have eight or more facets between them. By Lemma~\ref{l:centroid}, two facets do not overlap when the distance between their centroids is greater than $\sqrt{3}$, which happens in every case.
\end{proof}

Buekenhout and Parker \cite{bupa} enumerate 110,912 ridge unfoldings of the 4-orthoplex.  The following guarantees all of them to be valid nets.

\begin{thm}
The 4-orthoplex is all-net.	
\end{thm}

\begin{proof}
If there were an unfolding of the 4-orthoplex that did not yield a net, then there would be a path between two of its overlapping facets. By Lemma~\ref{l:8more}, those facets must be separated by fewer than eight intervening facets along the path, corresponding to a valid list $\lst$ whose length is at most eight.  By Lemma~\ref{l:8less}, that list can be extended to one whose length is exactly eight. As described in Lemma~\ref{l:eight}, none of the unfolds generated by these lists exhibit overlap. 
\end{proof}

\subsection{Higher dimensions}

Although the $n$-cube is all-net \cite{ddrw}, it is surprising that its dual does not inherit this property:

\begin{thm}
For each $n>4$, the $n$-orthoplex is not all-net.
\end{thm}

\begin{proof}
We first consider case-by-case analysis of dimensions 5 through 9 using centroid arguments.
Consider the valid list 
$$\langle 1, 2, 1, 3, 1, 2, 1, 4, 1, 2, 3, 5, 1, 2, 3, 2, 5, 3, 2, 1, 5, 4, 3, 4, 2, 4, 1, 2, 3, 2, 1\rangle$$
that captures the path unfolding of all 32 facets of the 5-orthoplex. The centroid of its last facet is approximately 
	\[ ( 0.22687086, 0.04417632, 0.36540937, 0.12942886, 0.23411458) \,,\]
which is a distance of approximately $0.24188305539$ from the centroid of the first facet, $$(0.2,0.2,0.2,0.2,0.2) \,.$$ 
By Lemma~\ref{l:centroid}, since the centroids are separated by less than $1/\sqrt{5}$, 
their facets overlap.
In dimension 6, there is an even shorter chain (16 facets) that fails the centroid test, given by 
    \[ \langle 1,2,3,1,4,5,4,3,5,4,1,3,2,1,4 \rangle. \]
In dimensions 7 and 8, the shortest chain (13 facets) that fails the centroid test is given by
    \[ \langle 1,2,3,4,1,5,3,5,4,3,2,1 \rangle. \]
For dimension 9, there is an even shorter chain (10 facets) given by $\langle 1,2,3,4,2,4,1,2,3 \rangle.$

For dimensions $n>9$, we focus on the unfolding corresponding to the list  $\langle 1,2,3,4,2,4,1,2,3 \rangle$ above but the centroid argument will not be utilized. Recall that the unfolding of the orthoplex occurs in the hyperplane $\sum x_i=1$. The first facet is the intersection of this hyperplane with the positive orthant. Let 
$$v= \Bigl< \ \frac{1}{n-1} \ ,\ 0 \ ,\ \frac{1}{n-1}, \ \ldots, \ \frac{1}{n-1} \ \Bigr> \,,$$ 
the midpoint of a ridge of the first facet. We show that its image under the unfolding, which is on a ridge of the tenth facet, has all positive coordinates. Therefore it is a point of intersection of the first and tenth facets. To do this, let $x=2/(n-1)$. Then each $M_i$ can be written in terms of $x$, and 
$$v= \Bigl< \ \frac{x}{2} \ ,\ 0 \ ,\ \frac{x}{2}, \ \ldots, \ \frac{x}{2} \ \Bigr> \,.$$ 
The matrix product 
$$M_1 \cdot M_2 \cdot M_3 \cdot M_4 \cdot M_2 \cdot M_4 \cdot M_1 \cdot M_2 \cdot M_3 \cdot v$$
 is then a vector  $\langle p_1(x), p_2(x), \ldots, p_n(x) \rangle$ whose entries are polynomials in $x$. Note that although the variable $x$ depends on $n$, the polynomials themselves are the same for all $n$. Using {\tt Sympy} \cite{sympy}, these were calculated to be
\begin{align*}
p_1(x) &=  0.5 x^9 - 3 x^8 - 7 x^7 - 8 x^6 - 5 x^5 - 1.5 x^4 + x^3 + x^2 + 0.5x \\
p_2(x) &= 0.5x^{10} + 3 x^9 + 6.5 x^8 + 5.5 x^7 + 0.5 x^6 - 1.5 x^5 - 0.5 x^4 + x^3 + x^2 \\
p_3(x) &= 0.5 x^{10} + 3.5 x^9 + 9.5 x^8 + 12 x^7 + 6 x^6 - 2 x^5 - 5.5 x^4 - 4 x^3 -  x^2 + 0.5x \\
p_4(x) &= 0.5 x^{10} + 3.5 x^9 + 10 x^8 + 14.5 x^7 + 10.5 x^6 + 2 x^5 - 2.5 x^4 - 1.5 x^3 + 0.5x 
\end{align*}
and for $i \ge 5$,
    \[ p_i(x) = 0.5 x^{10} + 3.5 x^9 + 10 x^8 + 15 x^7 + 13 x^6 + 6.5 x^5 + x^4 - 0.5 x^3 + 0.5x. \]
The lowest degree term, which determines the behavior near zero, is positive in each.  Thus, for sufficiently small $x$, each is positive. In fact, their graphs show that they are all positive when $0 \le x \le 0.2278$. Since $x=2/(n-1)$, this corresponds to all $n>9$. 
\end{proof}

\begin{rem}
The shortest chains that fail the centroid test in dimension 5 have length 20. One is given by the list $\langle 1,2,3,4,2,1,5,4,2,4,5,4,2,1,5,4,3,1,5 \rangle$.
It is possible that there are shorter chains that overlap even though their centroids are not sufficiently close to guarantee it.
\end{rem}

\begin{rem}
For dimensions $n>9$, the list  $\langle 1,2,3,4,2,4,1,2,3 \rangle$ provides a 10-length chain of facets that overlaps.  It would be interesting to discover if this is the shortest length chain with this property.
\end{rem}    

\subsection{Closing}
Horiyama and Shoji \cite{hosh} showed that the five regular polytopes in 3D (the Platonic solids) are all-net. Three classes of regular polytopes exist for all dimensions: the  $n$-simplex, $n$-cube, and $n$-orthoplex.  The $n$-simplex and $n$-cube are all-net, but the $n$-orthoplex fails (except for $n=4$).  There are only three additional regular polytopes whose all-net property has not been studied, all of which are four-dimensional: the 24-cell, 120-cell, and 600-cell.  The number of distinct unfoldings of these three polytopes are enumerated in \cite{bupa}:
\begin{align*}
{\text{24-cell}} & \ : \ \ 6 \ (2^{19} \cdot 5688888889 \ + \ 347) \\
{\text{120-cell}} & \ : \ \ 2^7 \cdot 5^2 \cdot 7^3 \ (2^{114} \cdot 3^{78} \cdot 5^{20} \cdot 7^{33} \ + \ 2^{47} \cdot 3^{18} \cdot 5^2 \cdot 7^{12} \cdot 53^{5} \cdot 2311^3 \ + \ 239^2 \cdot 3931^2) \\
{\text{600-cell}} & \ : \ \ 2^{188} \cdot 3^{102} \cdot 5^{20} \cdot 7^{36} \cdot 11^{48} \cdot 23^{48} \cdot 29^{30}
\end{align*}
Unlike the other three regular polytopes of this dimension (4-simplex, 4-cube, 4-orthoplex), these enormous unfolding numbers  point us towards the following claim:

\begin{conj} 
The 24-cell, the 120-cell, and the 600-cell fail to be all-net.
\end{conj}

Easing the condition of regularity for 3D polyhedra move us away from Platonic solids to the Archimedean solids.  However, it is known that several of these polyhedra (such as the snub dodecahedron,  truncated icosahedron,  truncated dodecahedron,  rhombicosidodecahedron, and  truncated icosidodecahedron) fail to be all-net \cite{hosh}.  It would be interesting to find conditions for polyhedra, and polytopes in general, that guarantee the all-net property.

%
%
\bibliographystyle{amsplain}

\end{document}